\documentclass[graybox]{svmult}

\usepackage{type1cm}        % activate if the above 3 fonts are
\usepackage{makeidx}         % allows index generation
\usepackage{graphicx}        % standard LaTeX graphics tool
\usepackage{multicol}        % used for the two-column index
\usepackage[bottom]{footmisc}% places footnotes at page bottom

\usepackage{newtxtext}       
\usepackage[varvw]{newtxmath}   

\makeindex
\begin{document}
\title*{From  ADONE's multi-hadron production to the $J/\Psi$ discovery}
\author{Mario Greco}
\institute{Mario Greco \at Department of Mathematics and Physics and INFN, University of Roma Tre, Rome, Italy, \email{mario.greco@roma3.infn.it}}

%Please use the 'starred' version of the \texttt{abstract} command for typesetting the text of the online abstracts (cf. source file of this chapter template \texttt{abstract}) and include them with the source files of your manuscript. Use the plain \texttt{abstract} command if the abstract is also to appear in the printed version of the book.}}
                                 	       
\maketitle	       

\abstract{The physics at Frascati in the years 60's - 70's is reviewed together with  the $J/\Psi$ discovery.}

\vskip 3 cm
	
        I arrived at the Frascati National Laboratory in January 1965,  joining the ADONE Group, and stayed there for 25 years. The first electron-positron collider AdA had been successfully constructed in one year only after the famous seminar given by Bruno Touschek on March 7th, 1960. Taken to Orsay to improve the injection rate obtainable with the linear
accelerator,  AdA completed its cycle there \cite{1}  and the last publication  \cite{2} was at the end of 1964. 
 On the contrary the ADONE adventure  took a much longer time. The first draft of ADONE and its physics programme dated November 1960 and was written in the Bruno Touschek's notebook,
 where he listed a number of physical processes which could be measured, including the production of a proton-antiproton pair.  The maximum total energy of the machine was then set to 3 GeV
 \cite{3}. ADONE's approval was in 1963 and  the first collisions were finally obtained  in 1969 with  the first generation experiments.
 
 The radiative corrections for ADONE experiments were the main problem Bruno Touschek had in his mind, because of their importance in the new electron-positron collider, and indeed they had been the main field of theoretical activity of the new theory group.  In addition the calculation of the double bremsstrahlung process as a monitor reaction for the luminosity was also performed
 using two different approaches \cite{4}. 
  In more detail the infrared corrections to be applied in an electron-positron collider experiment were first obtained with the help of the Bloch-Nordsieck theorem, using a statistical approach to define the probability for the four-momentum to be carried away by the electromagnetic radiation \cite{5}. 
 Alternatively, from a field theoretical point of view, a new finite S-Matrix was defined using a realistic definition of initial and final states, by "dressing" the charged particle's states with a phase containing the electromagnetic operator in the exponential, in order to create an undetermined number of soft photons with a Bloch-Nordsiek spectrum. The new S-Matrix was explicitly shown to be equivalent to all orders in $\alpha$ 
to the conventional perturbative result \cite{6}. 
In other words this approach corresponds to the introduction of the idea of coherent states in QED. It is extraordinary that both approaches led exactly to the same result for the soft radiative effect, namely that the observed cross section can be written as
\begin{equation}
d\sigma=\frac{1}{\gamma^\beta \Gamma(1+\beta)} \left( \frac{\Delta \omega}{E} \right)^\beta d\sigma_E \nonumber
 \end{equation}  
   
  where 2E is the total c.m. energy,    $(\Delta \omega/E)$
is the relative energy resolution of the experiment,  $\gamma $ 
is the Euler constant, $d\sigma_E$ 
differs from the lowest cross section $d\sigma_0$ 
 by finite terms of $\mathcal{O(\alpha)}$, and $\beta$ 
 is the famous Bond-factor, so named by Bruno because its numerical value at ADONE was 0.07, and more generally $ \beta= \frac{4}{\pi} [\ln (2E/m) -1/2]$. 	
  
The coherent states approach played a major role later in the description of the radiative effects in case of the production of the $J/\Psi$ 
and of the Z boson, as we'll discuss later. Also that was first extended to QCD in the late '70 \cite{7}
and studied further \cite{8,9}.  
Many and important QCD results concerning exponentiation, resummation formulae, K-factors, transverse momentum distributions of DY pairs, W/Z and H production, have their roots in Bruno Touschek ideas on the exponentiation and resummation formulae in QED. We give here a reference list 
\cite{10} of the papers that were written later in the Frascati-Rome area and certainly inspired by his ideas.

 Let's discuss now the theoretical framework and the expectations concerning ADONE and the experimental results. At the time, the Vector Meson Dominance (VMD) model of J.J. Sakurai \cite{11} 
was quite successful in describing the e.m. interaction  of hadrons as being mediated by the vector mesons $\rho, \omega$ and $\phi$.
That led T.D. Lee, N. Kroll and B. Zumino to try to give a field theoretical approach to VMD \cite{12}. 
 In this framework the total hadronic annihilation cross section was expected to behave at large $s$ as 
\begin{equation}
 \sigma(s)=\left( \frac{1}{s} \right)^2 \nonumber
\end{equation}
			However departures from the simple VMD model were observed in some radiative decays of mesons and the possible existence of new vector mesons was suggested by A. Bramon and myself  \cite{13}, 
 as also predicted by dual resonance models and the Veneziano model \cite{14}.  
 On the other hand, the results of Deep Inelastic Scattering (DIS) experiments at SLAC, with the idea of Bjorken scaling and the Feynman parton model were naturally leading to 
\begin{equation}
\sigma(s)=\frac{1}{s}\nonumber
\end{equation}			
and indeed N. Cabibbo, M. Testa and G. Parisi \cite{15} 
 suggested that the ratio R of the hadronic to the point-like cross section would asymptotically behave as
 \begin{equation}
 R=\frac{\sigma_{had}(s)}{\sigma_{\mu \mu}}\rightarrow \sum_i Q_i^2
 \nonumber
\end{equation}             
where the sum extends to all spin 1/2 elementary constituents, neglecting scalars. 

As it's well known, the results of all experiments, namely the MEA Group \cite{16}, 
the $\gamma \gamma $  Group \cite{17},  the  $\mu \pi$  
Group \cite{18}, and the Bologna-CERN-Frascati Collaboration \cite{19} 
showed a clear evidence of a large multihadron production with $R \approx 2$, pointing to the coloured quark model. On the other hand they also indicated evidence for a new vector meson $\rho' (1.6)$ with a dominant decay in four charged pions, which had been suggested by A. Bramon and myself  \cite{20}.
The experimental data are shown in Fig.1, 
taken from a review paper of C. Bernardini and L. Paoluzi \cite{21}. 
  %\begin{figure}\centering  \includegraphicx[]{}  \caption{}\label{}  \end{figure}
\begin{figure}[t]
\centering%%to decide latr \sidecaption
% For example, with the graphicx style use
\includegraphics[scale=0.2]{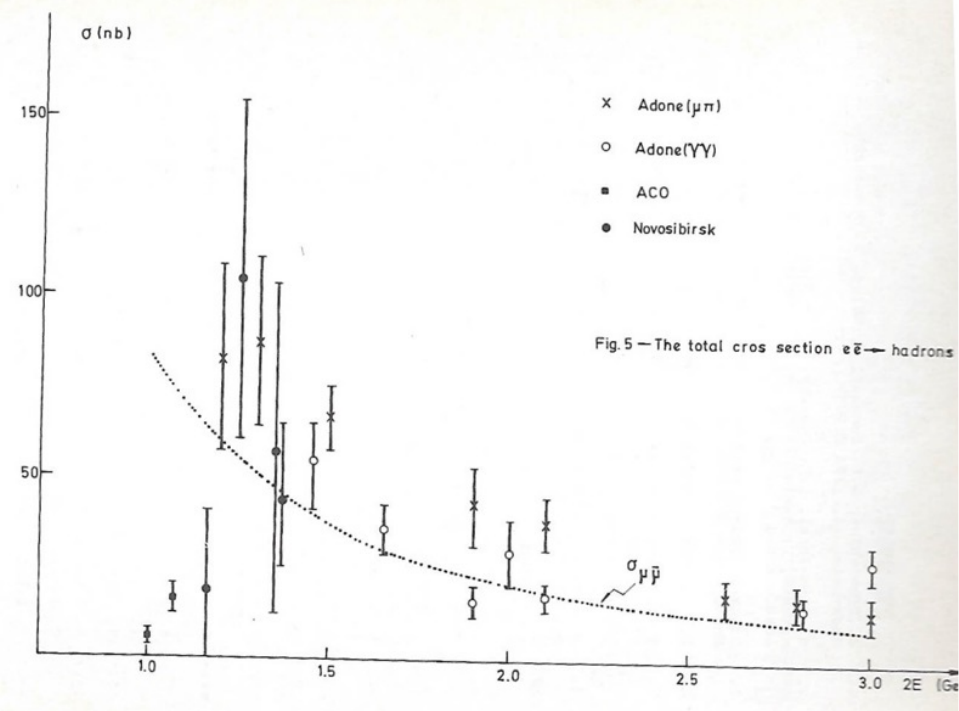}
\includegraphics[scale=0.2]{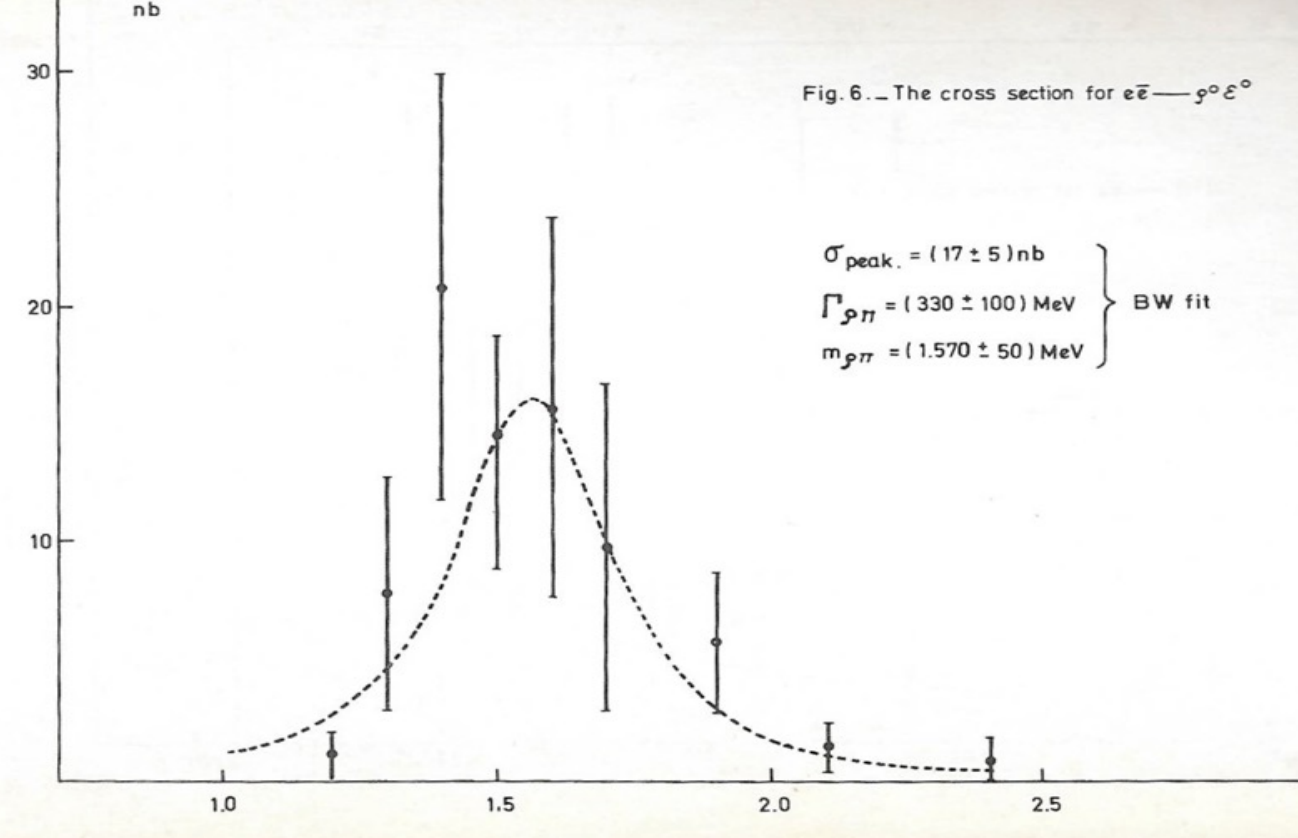}
%
% If no graphics program available, insert a blank space i.e. use
%\picplace{5cm}{2cm} % Give the correct figure height and width in cm
%
\caption{ADONE experimental results, from ref. [21].}
\label{fig:1-2}       % Give a unique label
\end{figure}

The ADONE results together with the request of scaling, both in DIS and e+e- annihilation, and the Veneziano's duality idea led us to propose a new scheme where the asymptotic scaling is reached through the low energy resonances mediating the asymptotic behaviour \cite{22}. 
 Thus the value of R is also connected to the low energy resonances's couplings, and in the 3 coloured quark model, led us to the prediction      $ R \approx 2.4$.  This scheme - named duality in $e^+e^-$ %e+e- 
 annihilation - was immediately shared by  J. J. Sakurai \cite{23}. 
 Later J. Bell and collaborators also studied a potential model where the bound states could be solved analytically and verified this idea of duality \cite{24}. 
 In addition a set of $e^+e^-$ %e+e- 
 duality sum rules was derived from the canonical trace anomaly of the energy momentum tensor by E. Etim and myself \cite{25},
 much earlier than the Russian sum rules of M. A. Shifman et al. \cite{26}. 
 The lowest order sum rule gives
\begin{equation} 
\int_{s_0}^{\bar s} ds {\left( Im \Pi(s)- \frac{\alpha R}{3}\right) }=0 
\nonumber
\end{equation}
where $Im \Pi (s)= s \sigma_{had}(s)/4\pi \alpha$
%     ImΠ (s) = s σhad(s) / 4πα,
and clearly it relates the asymptotic value of $R$ to the low energy behaviour. One has to stress here that QCD wasn't there yet at that time. The average value of $R$ in particular, in the ADONE region, is about 2.4 as it was also confirmed by the SPEAR data at the c.m. energy just below the $J/\Psi$. On the other hand, for larger c.m. energies, the SPEAR data  also showed an increasing behaviour of $R$ suggesting the presence of a new component in $Im \Pi (s)$ with threshold at about 3 GeV.  That was our conclusion in ref. \cite{25}, which dates a few weeks before the $J/\Psi$  discovery.

Let's consider in detail now the  $J/\Psi$  discovery, or what was called the November Revolution, from a Frascati point of perspective. As it's well known, on November 11th 1974, B. Richter and S.C.C. Ting jointly announced in Stanford the discovery of the. $J/\Psi$  %J/Ψ
 both at SLAC and at Brookhaven \cite{27,28}. 
  I had the terrific chance of arriving at SLAC the day after, with an invitation by Sid Drell to give a seminar on our duality works, on the way for a visit of a few weeks to Mexico City.  Sid had been on a sabbatical leave the year before at Frascati and Rome, so we knew each other pretty well.There was a great excitement in the theory discussion room and once I was informed of the details of the discovery I realized immediately that the  $J/\Psi$ %J/Ψ
  could be seen possibly also at ADONE. I asked Sid to let me call Frascati, and from his confidential office - he was scientific advisor of the President of United States - I gave to Giorgio Bellettini, the director of the Laboratory, the exact position of the $J/\Psi$. 
   The night after, the resonance was also observed at Frascati. Giorgio Bellettini communicated the results to the Phys. Rev. Letters over the telephone and the paper was published \cite{29} 
   in the same issue of the American results. 
  
\begin{figure}[t]
  \centering
%%to decide latr \sidecaption
% For example, with the graphicx style use
\includegraphics[scale=.4]{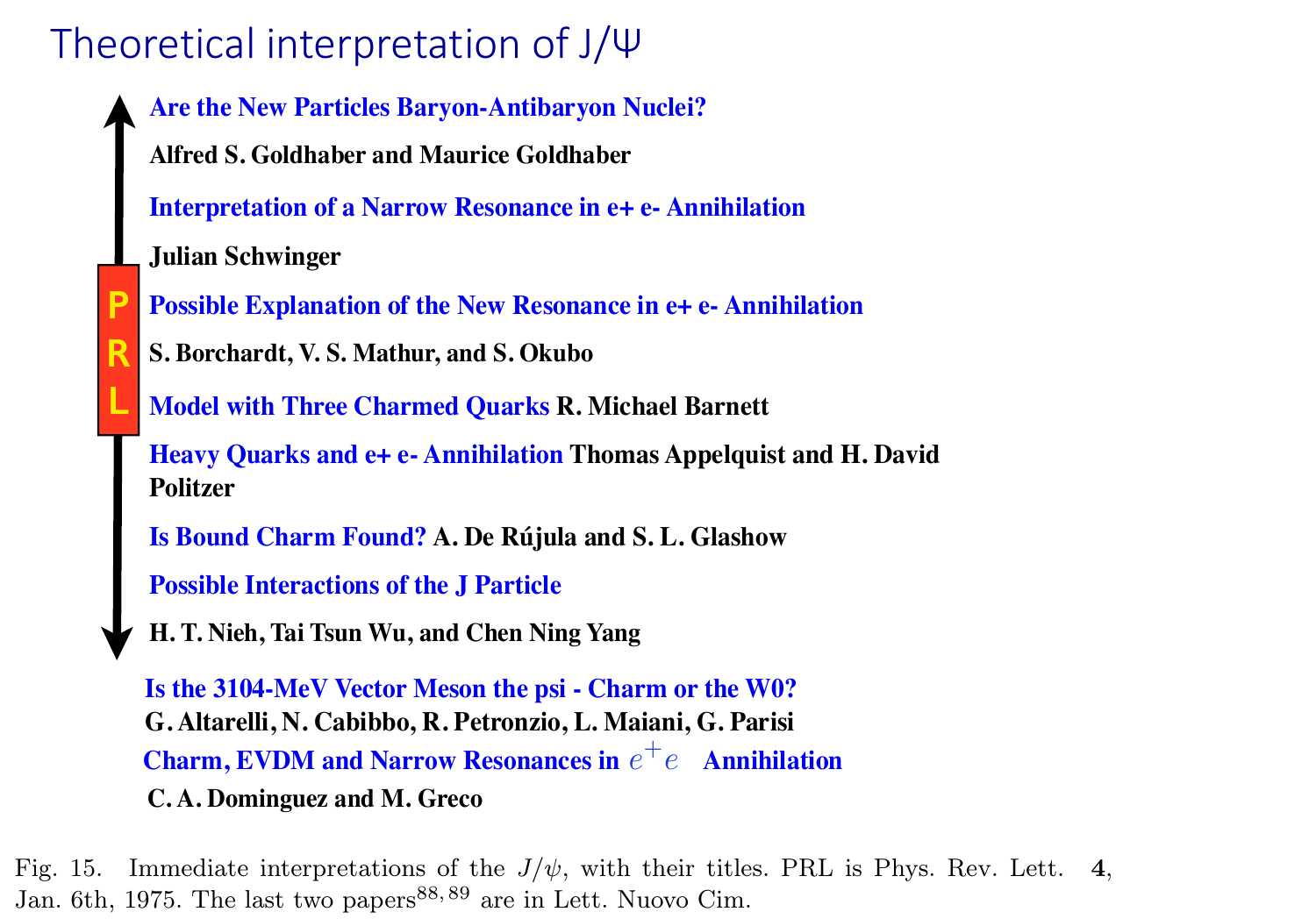}
%
% If no graphics program available, insert a blank space i.e. use
%\picplace{5cm}{2cm} % Give the correct figure height and width in cm
%
\caption{The immediate interpretation of $J/\Psi$ from ref. [30]. PRL is Phys. Rev. Letts. 34, Jan. 16th, 1975.}
\label{fig:3}       % Give a unique label
\end{figure}

As far as the theoretical interpretation of the $J/\Psi$ is concerned,  
hundreds of papers had been published on the argument, as it's well known. In a recent review article on this subject, Alvaro De Rujula has reported \cite{30} the papers published on the first issue of Phys. Rev. Letters after the discovery, as shown in Fig.\ref{fig:3}.
 %[5].

Of course only two of them, both from Harvard, had the right interpretation: by T. Applequist and H.D. Politzer \cite{31}, 
who related the reason for the very narrow width of the  $J/\Psi$ 
to the asymptotic freedom of QCD just discovered, and by A. De Rujula and S.L. Glashow \cite{32} 
 because of the GIM mechanism and charm suggested earlier \cite{33}. 
 On the other hand Alvaro is also quoting two papers published on Lett. Nuovo Cimento, by G. Altarelli, N. Cabibbo, L. Maiani, R. Petronzio and G. Parisi \cite{34}, 
 who had the wrong interpretation in favour of the weak boson, and C. Dominguez and myself  \cite{35}, 
  written in Mexico City a few days after I had left Stanford, who also had the right interpretation in favour of charm. 
  Indeed I had taken with me the detailed data of the  $J/\Psi$ and as soon as I found the news on a local newspaper of the subsequent discovery of the  $\Psi'$,   
by using the duality ideas discussed above, we arrived at the conclusion that the new series of resonances was indeed composed by $c-{\bar c}$ pairs, and  the value of R would reach  about 3.7 with the new charm contribution.  On the other hand the limits of our paper were both a very naive assumption on the $\Psi's$ mass spectrum and no understanding of the smallness of the 
 $J/\Psi$ width.
 
Forty years later, in December 2013, after Sam Ting  had mentioned it  in his "Bruno Touschek Memorial Lecture" in Frascati, I discovered in the archives
at the CERN library  that our preprint was dated November 18, preceding by a few days those of Applequist and Politzer and of De Rujula and Glashow.

\begin{figure}[t]
 \centering
%%to decide latr \sidecaption
% For example, with the graphicx style use
\includegraphics[scale=.4]{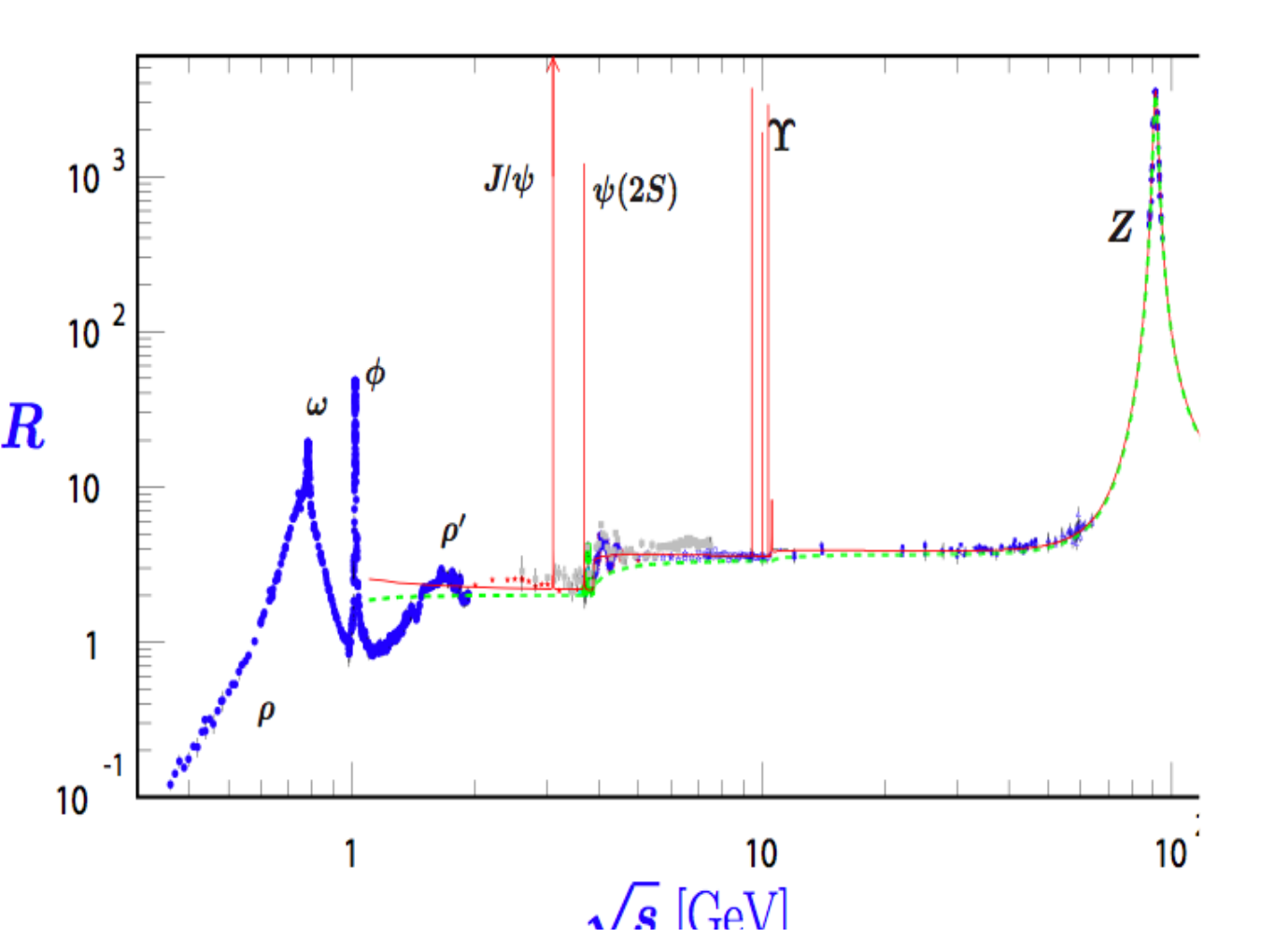}
%
% If no graphics program available, insert a blank space i.e. use
%\picplace{5cm}{2cm} % Give the correct figure height and width in cm
%
\caption{The ratio $R = \sigma_{had}(s)/\sigma_{\mu\mu}(s)$ as a function of $\sqrt{s}$, from Particle Data Group.}
\label{fig:4}       % Give a unique label
\end{figure}

Now, fifty years later, by comparing  the value of R from the Particle Data Group with all the experimental information, as shown in Fig.\ref{fig:4}, 
with the theoretical prediction of QCD with $\mathcal{O}(\alpha_s) $, $\mathcal{O}(\alpha_s^2) $ and $\mathcal{O}(\alpha_s^3 )$
corrections included - as indicated by the continuous red line - one easily concludes that our duality predictions were  well satisfied, and indeed  in very good agreement with QCD.

\begin{figure}[t]
 \centering
%%to decide latr \sidecaption
% For example, with the graphicx style use
\includegraphics[scale=.4]{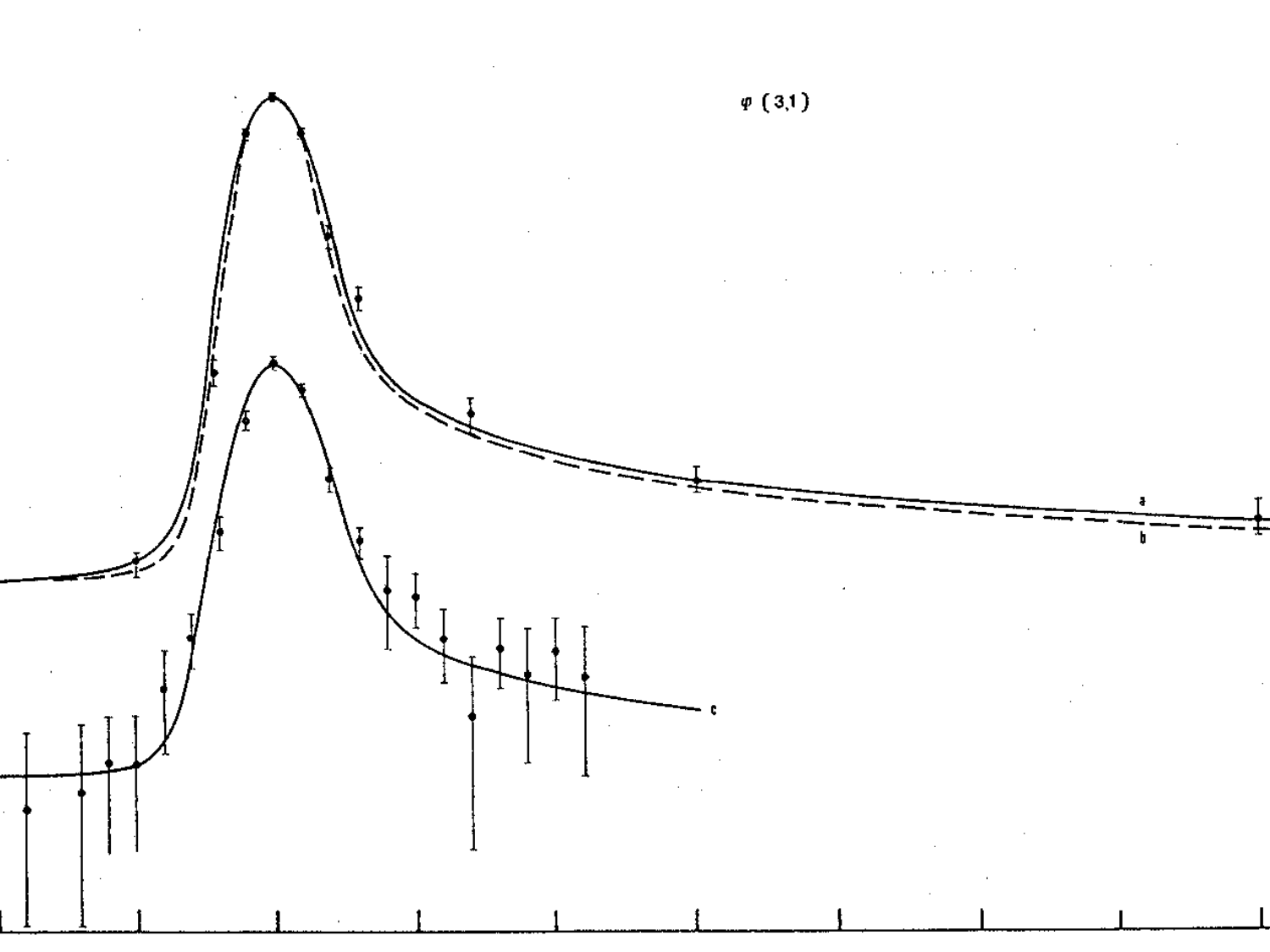}
%
% If no graphics program available, insert a blank space i.e. use
%\picplace{5cm}{2cm} % Give the correct figure height and width in cm
%
\caption{Experimental results for $J/\Psi$  production in $e^+e^-$ annihilation. The data are from SPEAR and ADONE (see text). The full lines refer to the theoretical analysis including radiative corrections of Ref.~\cite{36}.}
%If the width of the figure is less than 7.8 cm use the \texttt{sidecapion} command to flush the caption on the left side of the page. If the figure is positioned at the top of the page, align the sidecaption with the top of the figure -- to achieve this you simply need to use the optional argument \texttt{[t]} with the \texttt{sidecaption} command}
\label{fig:5}       % Give a unique label
\end{figure}

The problem of the radiative corrections to the  $J/\Psi$ 
 line-shape, because of the very narrow width involved, showed the crucial role played by the theoretical ideas of the early times on the infrared behaviour of QED, namely the exponentiation results and the approach of the coherent states. The detailed analysis by G. Pancheri, Y. Srivastava and myself  \cite{36}, 
 showed that the main infrared correction factor was of the type
\begin{equation}
C_{infra} \approx \left(\frac{\Gamma}{M}\right)^\beta\nonumber
\end{equation}	                                                                   			
%C infra   ≈  (Γ/M) β           				  									  
where $\Gamma$ and $M$ are the width and the mass of the resonance respectively, and $\beta$ 
 the Bond-factor.  The detailed result of this analysis, compared with the SLAC and Frascati data, is shown in Fig.~\ref{fig:5}. 
 When we showed our result to Bruno Touschek, he immediately commented that the experimental errors of the Frascati data had been clearly overestimated. 
  I should also add that the SLAC analysis of their data had been based on a paper by D.R. Yennie \cite{37} 
 that contained a wrong dependence on the width $\Gamma$ 
 %Γ
  and the parameter $\sigma$
  of the Gaussian energy distribution of the beams, with a resulting difference with respect to our analysis on the leptonic width of the $J/\Psi$. %J/Ψ.
   It was only in 1987, in the occasion of the first La Thuile meeting, that I convinced Burt Richter to update the SLAC radiative corrections codes with the right formulae, in perspective of the coming data on the Z boson physics at SLC. Indeed the re-analysis of all charm data at SLAC caused a change of many properties of the charm particles in the Particle Data Group in 1988, including 
the leptonic width of the $J/\Psi$, in better agreement with our estimate.

The above treatment of the radiative corrections for the $J/\Psi$ 
 production was extended a few years later to study the radiative effects in the case of Z production at LEP/LHC \cite{38}. 
 Our work was the first study to all orders in the infrared corrections, with a complete evaluation of all finite terms of  $\mathcal{O(\alpha)}$
  and at the base of the later analyses of all experiments. Very recently, within the general discussion on the possibility of constructing a muon collider Higgs factory to study with great care on resonance the properties of the H, the line-shape has been studied \cite{39}
  in the same way, as in the old times. As a result we have shown that the radiative effects put very stringent bounds on the energy spread of the beams, and make this project very tough. 

To conclude, from AdA/ADONE to LEP/LHC and the future  colliders, the seminal idea of Bruno Touschek has contributed with so many discoveries to  the progress of the Standard Model. This certainly constitutes his main legacy. In addition some of  his suggestions  have also strongly contributed to the  precision assessment of the theory.  Finally duality  ideas have been a very powerful tool for the interpretation of e+e-  colliding beams results and in particle phenomenology before the advent of QCD.

\end{document}